\PassOptionsToPackage{table}{xcolor}

\documentclass[conference]{IEEEtran}
\usepackage[utf8]{inputenc}
\usepackage{csquotes}
\usepackage{xcolor}
\usepackage{algorithm}
\usepackage{algpseudocode}
\usepackage{xspace}
\usepackage{listings}
\usepackage{color}
\usepackage{multirow}
\usepackage{dblfloatfix}
\usepackage{lscape} 
\usepackage{xparse}
\usepackage{placeins}
\usepackage{caption} 
\usepackage{subcaption}
\usepackage{enumitem}
\usepackage{array, makecell}
\usepackage{wrapfig}
\usepackage{colortbl}
\usepackage{tabularx} 
\usepackage{booktabs}
\usepackage{courier}
\usepackage{ragged2e}
\usepackage{amsthm}
\usepackage{amsmath,amssymb,amsfonts}
\usepackage{hyperref}
\usepackage{comment}
\usepackage{pdflscape}
\usepackage{afterpage}
\usepackage{fontawesome}
\usepackage{pifont}
\usepackage{lscape}
\usepackage{svg}
\usepackage{pgfplots}
\usepackage[normalem]{ulem}

\setlist[itemize]{noitemsep, topsep=4pt}
\definecolor{lightgray}{HTML}{f6f6f6}
\definecolor{darkgray}{rgb}{.4,.4,.4}
\definecolor{darkblue}{HTML}{1b4db3}
\definecolor{brickred}{HTML}{b04f4f}
\definecolor{purple}{rgb}{0.65, 0.12, 0.82}
\definecolor{diffadd}{HTML}{288f26}
\definecolor{diffrmbg}{HTML}{ffebe9}
\definecolor{diffaddbg}{HTML}{e6ffeb}
\definecolor{diffremove}{HTML}{de4f54}
\definecolor{carrotorange}{rgb}{0.8, 0.33, 0.0}
\definecolor{highlight}{HTML}{fefbc2}
\lstdefinelanguage{JavaScript}{
  keywords={typeof, new, true, false, catch, function, return, null, catch, switch, var, const, let, extends, if, in, while, do, else, case, break, async, await,of,
  expect, field, toBeTruthy, toHaveLengthCondition, toBeAlphabetical, not, toBeEqual, fill, submit_form, assert, toBeNumerical
  },
  keywordstyle=\color{darkblue}\bfseries,
  ndkeywords={class, export, boolean, throw, implements, import, this, setTimeout},
  ndkeywordstyle=\color{brickred}\bfseries,
  identifierstyle=\color{black},
  sensitive=false,
  comment=[l]{//},
  morecomment=[f][\color{diffadd}\bfseries]{+\ },
  morecomment=[s]{/*}{*/},
  morecomment=[f][\color{diffremove}\bfseries]{- },
  commentstyle=\color{violet}\ttfamily,
  stringstyle=\color{carrotorange}\ttfamily,
  morestring=[b]',
  morestring=[b]"
}

\theoremstyle{definition}
\newtheorem{definition}{Definition}
\newcommand{\header}[1]{\par\smallskip\noindent\textbf{#1.}}

\def\BibTeX{{\rm B\kern-.05em{\sc i\kern-.025em b}\kern-.08em
    T\kern-.1667em\lower.7ex\hbox{E}\kern-.125emX}}
    
\newboolean{showcomments}
\setboolean{showcomments}{true}
\ifthenelse{\boolean{showcomments}}
{
	\definecolor{myyellow}{RGB}{255, 228, 26}
	\definecolor{myblue}{RGB}{50, 50, 220}
	\newcommand{\nb}[2]{
		{\sf
			\fcolorbox{myyellow}{yellow}{\scriptsize\textbf{#1}}%
			$\blacktriangleright$%
			{\color{myblue}\fontsize{7pt}{8pt}\selectfont\textbf{#2}}%
		}%
	}
}
{
	\newcommand{\nb}[2]{}
}

\newboolean{showmaybe}
\setboolean{showmaybe}{true}
\ifthenelse{\boolean{showmaybe}}
{
	\definecolor{myyellow}{RGB}{255, 228, 26}
	\definecolor{myred}{RGB}{184, 37, 95}
	\newcommand{\maybe}[1]{
		{\sf
			\fcolorbox{myyellow}{yellow}{\scriptsize\textbf{Maybe}}%
			$\blacktriangleright$%
			{\color{myred}\fontsize{7pt}{8pt}\selectfont\textbf{#1}}%
		}%
	}
}
{
	\newcommand{\maybe}[1]{}
}

\newcommand{\code}[1]{{\small\ttfamily\texttt{#1}}}

\algnewcommand\algorithmicforeach{\textbf{foreach}}
\algdef{S}[FOR]{ForEach}[1]{\algorithmicforeach\ #1\ \algorithmicdo}
\newcolumntype{Y}{>{\centering\arraybackslash}X}

\makeatletter
\DeclareRobustCommand{\change}{%
  \@bsphack
  \leavevmode
  \color{blue}
  \@esphack
}
\DeclareRobustCommand{\stopchange}{%
  \@bsphack
  \normalcolor
  \@esphack
}
\makeatother

\makeatletter
\newcommand{\linebreakand}{
  \end{@IEEEauthorhalign}
  \hfill\mbox{}\par
  \mbox{}\hfill\begin{@IEEEauthorhalign}
}
\makeatother

\author{
    \IEEEauthorblockN{Parsa Alian}
    \IEEEauthorblockA{
        \textit{University of British Columbia}\\
        Vancouver, Canada \\
        palian@ece.ubc.ca
    }
    \and
    \IEEEauthorblockN{Noor Nashid}
    \IEEEauthorblockA{
        \textit{University of British Columbia}\\
        Vancouver, Canada \\
        nashid@ece.ubc.ca
    }
    \and
    \IEEEauthorblockN{Mobina Shahbandeh}
    \IEEEauthorblockA{\textit{University of British Columbia}\\
        Vancouver, Canada \\
        mobinashb@ece.ubc.ca
    }
    \linebreakand
    \IEEEauthorblockN{Taha Shabani}
    \IEEEauthorblockA{\textit{University of British Columbia}\\
        Vancouver, Canada \\
        taha.shabani@ece.ubc.ca
    }
    \and
    \IEEEauthorblockN{Ali Mesbah}
    \IEEEauthorblockA{\textit{University of British Columbia}\\
        Vancouver, Canada \\
        amesbah@ece.ubc.ca
    }
}

\begin{document}

\title{Feature-Driven End-To-End Test Generation}


\maketitle


\newcommand{\toolname}{\textsc{AutoE2E}\xspace}
\newcommand{\benchname}{\textsc{E2EBench}\xspace}
\newcommand{\crawljax}{\textsc{Crawljax}\xspace}
\newcommand{\autogpt}{\textsc{AutoGPT}\xspace}
\newcommand{\opendevin}{\textsc{OpenDevin}\xspace}
\newcommand{\webcanvas}{\textsc{WebCanvas}\xspace}
\newcommand{\browsergym}{\textsc{BrowserGym}\xspace}

\newcommand{\funcdb}{\textsc{FD}\xspace}
\newcommand{\actionfuncdb}{\textsc{AFD}\xspace}

\newcommand{\gpt}{\textsc{GPT-4o}\xspace}
\newcommand{\claude}{\textsc{Claude 3}\xspace}
\newcommand{\sonnet}{\textsc{Claude 3.5 Sonnet}\xspace}

\newcommand{\subjectcount}{8\xspace}

\newcommand{\crawljaxcoverage}{12.0\%\xspace}
\newcommand{\webcanvascoverage}{0\%\xspace}
\newcommand{\browsergymcoverage}{9.5\%\xspace}
\newcommand{\autogptcoverage}{6.1\%\xspace}
\newcommand{\opendevincoverage}{7.9\%\xspace}
\newcommand{\ourscoverage}{\textbf{79}\%\xspace}
\newcommand{\totalfeaturecoverage}{72\%\xspace}

\newcommand{\crawljaxf}{0.62\xspace}
\newcommand{\browsergymf}{0.11\xspace}
\newcommand{\autogptf}{0.08\xspace}
\newcommand{\opendevinf}{0.08\xspace}
\newcommand{\oursf}{0.62\xspace}

\newcommand{\oursimprovement}{\textbf{558}\%\xspace}
\newcommand{\oursimprovementagent}{\textbf{731}\%\xspace}

\newcommand{\ourschain}{3.8\xspace}
\newcommand{\crawljaxchain}{2.9\xspace}
\newcommand{\browsergymchain}{1.4\xspace}
\newcommand{\autogptchain}{1.2\xspace}
\newcommand{\opendevinchain}{1.7\xspace}
\newcommand{\totalchain}{3.4\xspace}

\begin{abstract}
End-to-end (E2E) testing is essential for ensuring web application quality. However, manual test creation is time-consuming, and current test generation techniques produce incoherent tests. In this paper, we present \toolname, a novel approach that leverages Large Language Models (LLMs) to automate the generation of semantically meaningful feature-driven E2E test cases for web applications. \toolname intelligently infers potential features within a web application and translates them into executable test scenarios. Furthermore, we address a critical gap in the research community by introducing \benchname, a new benchmark for automatically assessing the feature coverage of E2E test suites. Our evaluation on \benchname demonstrates that \toolname achieves an average feature coverage of \ourscoverage, outperforming the best baseline by \oursimprovement, highlighting its effectiveness in generating high-quality, comprehensive test cases.
\end{abstract}

\maketitle

\section{Introduction}
\label{sec:introduction}
\begin{figure*}[h]
    \centering
    \begin{subfigure}[b]{0.24\textwidth}
         \centering
         \includegraphics[width=0.985\textwidth]{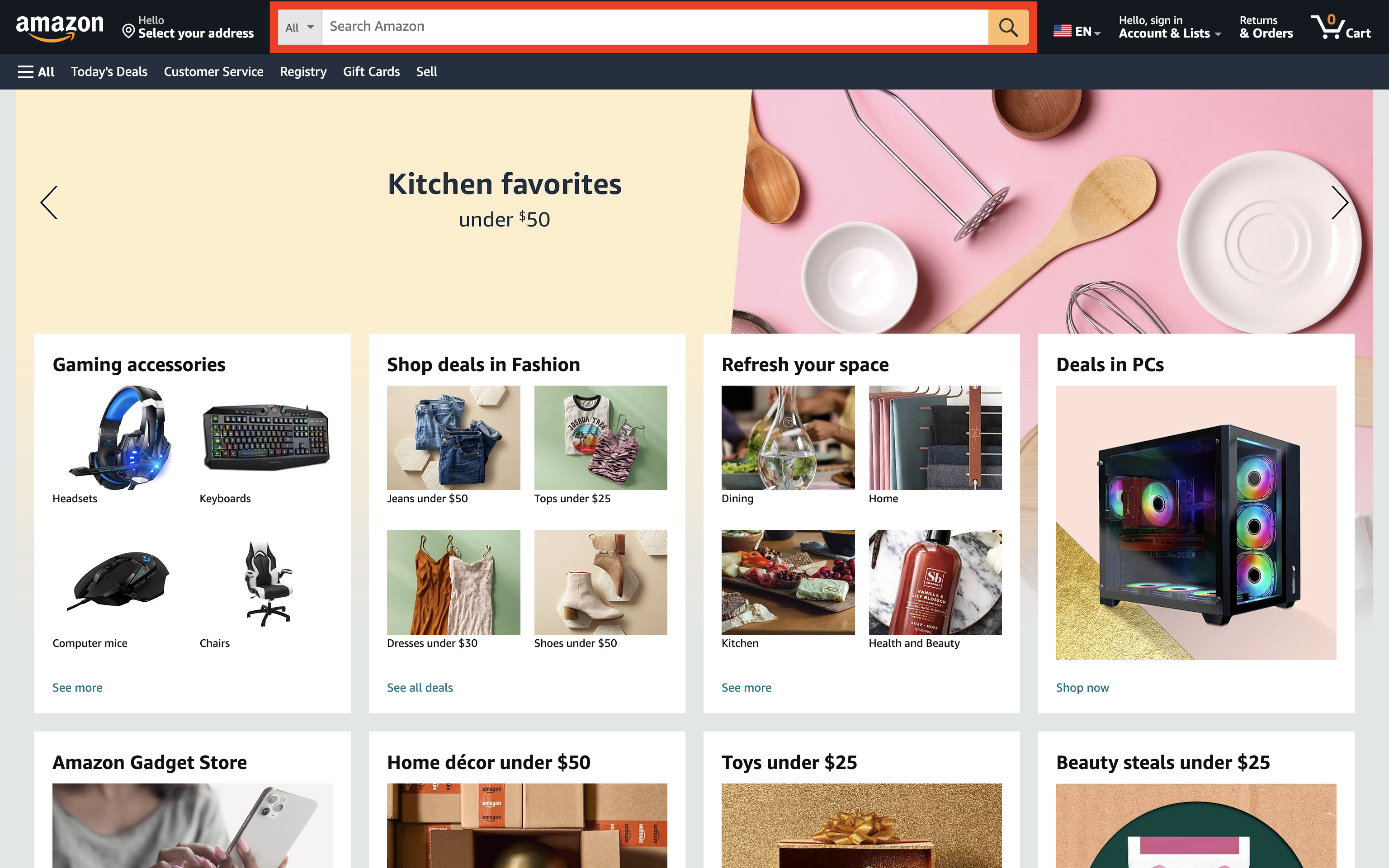}
         \caption{Select search bar}
         \label{fig:amazon-page-1}
     \end{subfigure}
     \begin{subfigure}[b]{0.24\textwidth}
         \centering
         \includegraphics[width=0.985\textwidth]{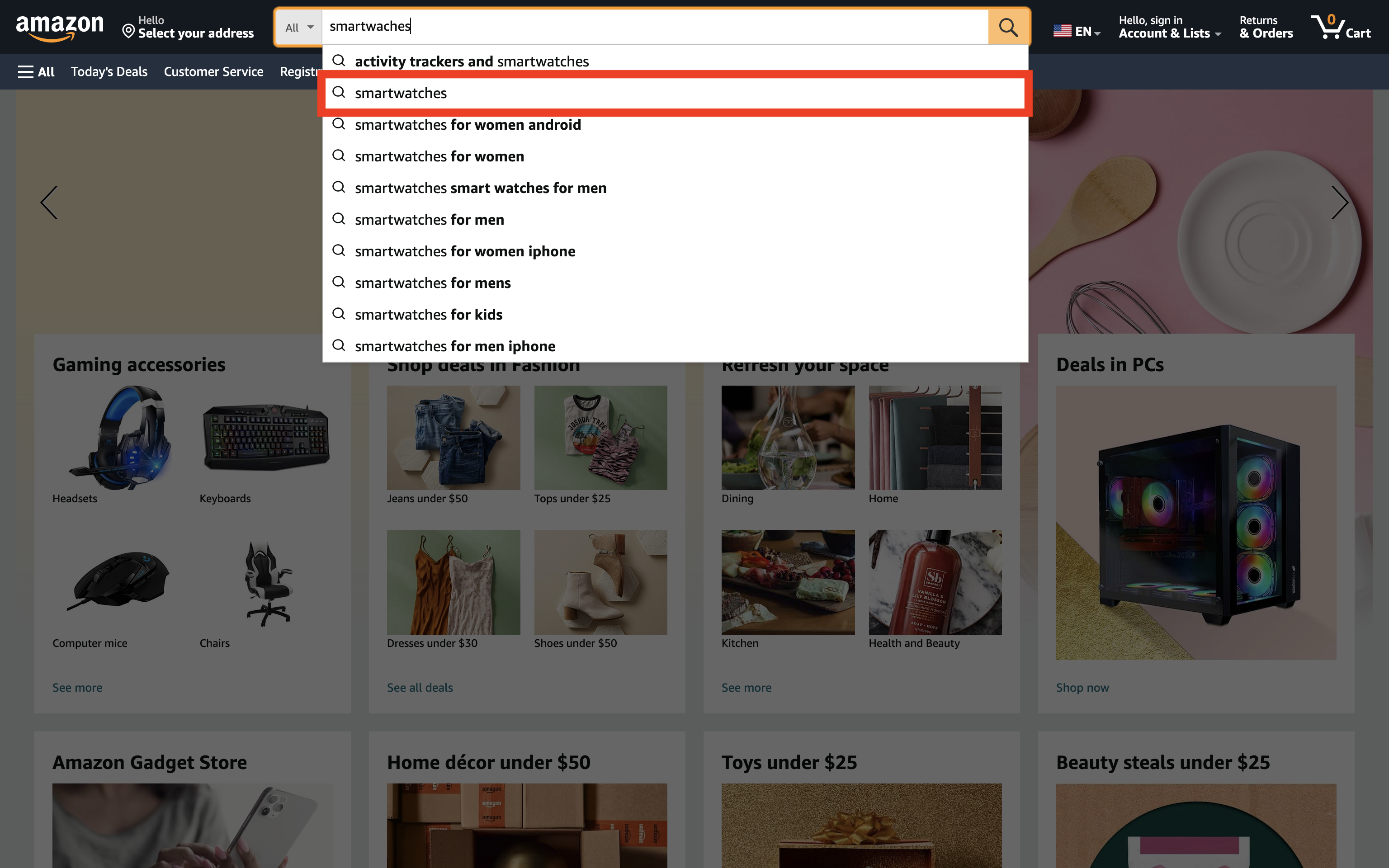}
         \caption{Search for a product}
         \label{fig:amazon-page-2}
     \end{subfigure}
     \begin{subfigure}[b]{0.24\textwidth}
         \centering
         \includegraphics[width=0.985\textwidth]{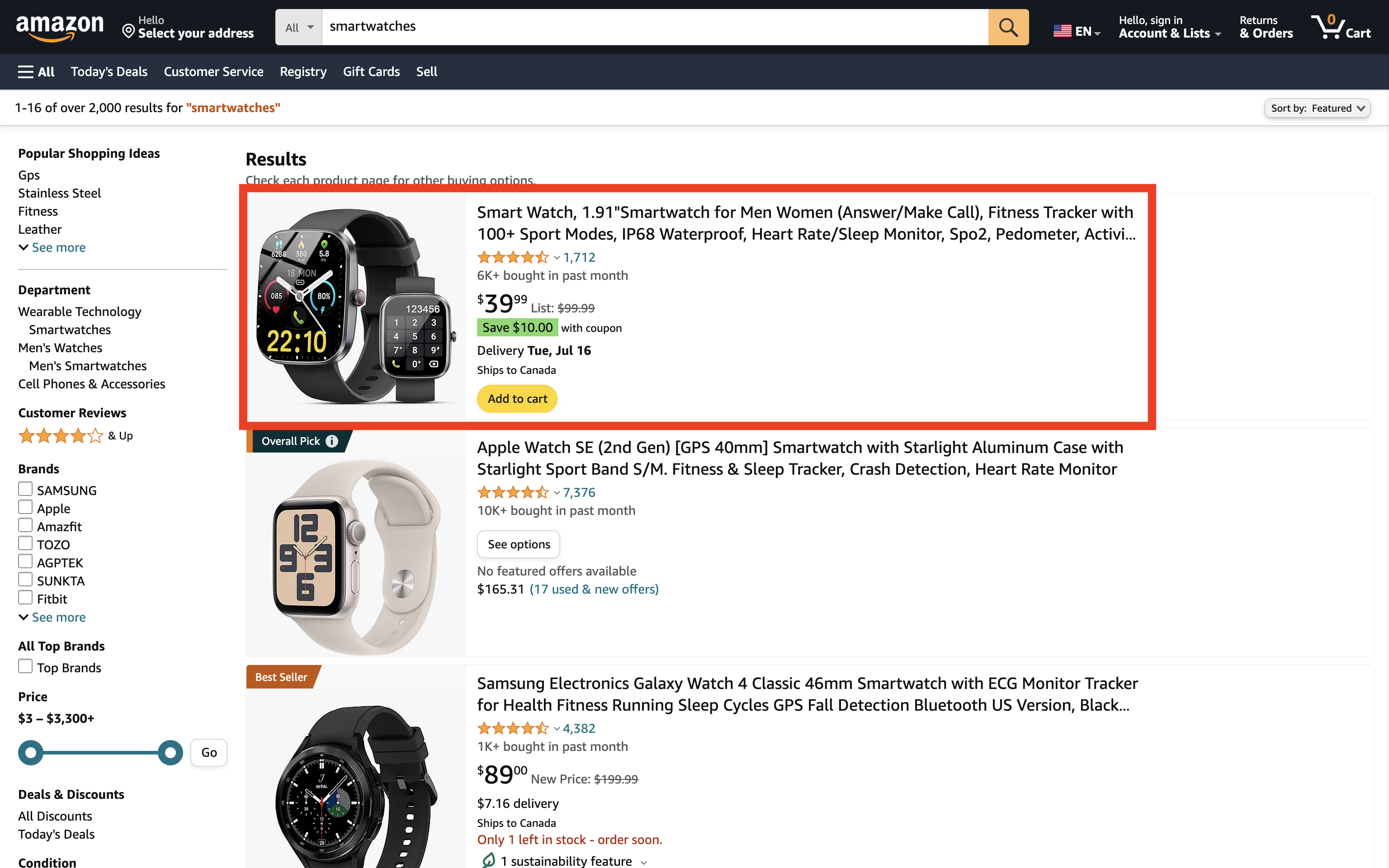}
         \caption{Select the product}
         \label{fig:amazon-page-3}
     \end{subfigure}
     \begin{subfigure}[b]{0.24\textwidth}
         \centering
         \includegraphics[width=0.985\textwidth]{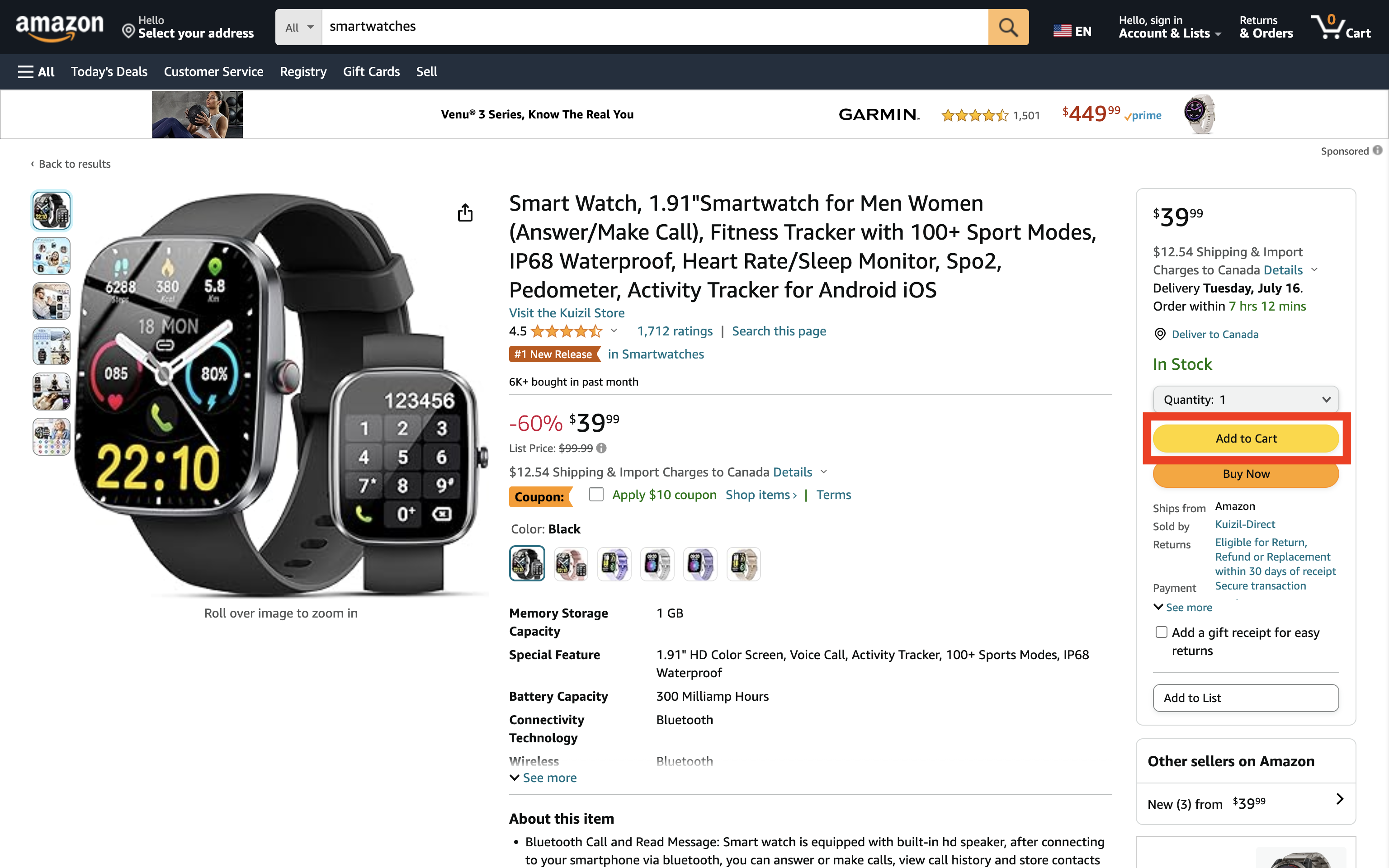}
         \caption{Add to cart}
         \label{fig:amazon-page-4}
     \end{subfigure}
    \caption{\textit{Add to Cart} feature on Amazon's web application}
    \label{fig:amazon-pages}
\end{figure*}

End-to-End (E2E) testing assesses whether various integrated components in an application work together correctly from the user interface (UI) to the back-end, by simulating real user interactions and verifying the application's functionality. In E2E, the application is tested as a whole, in its entirety, and from the perspective of the end-user~\cite{2019-ricca-Advances}.

The predominant method of creating E2E tests has relied heavily on human intervention, with developers manually assessing application features and using frameworks such as Selenium \cite{selenium} to script the user scenarios. Efforts to automate E2E test generation have explored reinforcement learning (RL) \cite{web-chang2023reinforcement} and model-based approaches~\cite{yueffective, web-fragmentsRahul, web-matteo-icst20}. More recently, the rise of large language models (LLMs) has spurred their application to various testing tasks. While LLMs have shown promise in generating unit tests for various applications \cite{xie2023chatunitest, kang2022large, lemieux2023codamosa, cedar, schafer2023adaptive} and mobile testing \cite{liu2024make, liu2024testing}, their application to web app E2E test generation remains an open area of research. Our recent work, FormNexus, has focused on automating web form testing~\cite{alian2024bridging}, highlighting the potential for LLMs to enhance E2E testing methodologies. 

In this paper, we formalize the notion of feature-driven E2E testing, including definitions of application features and a new metric called feature coverage for assessing E2E tests. Then, we propose \toolname, the first technique designed to generate feature-driven E2E tests autonomously.
Our approach is centered on the ability to automatically infer features embedded within the web application and translate them into a sequence of user actions that form an E2E test scenario. We approximate potential features and employ a novel probabilistic scoring method that deduces the likelihood of a feature's existence based on its observed frequency within the web application.
In our probabilistic approximation, we leverage LLMs, such as \gpt \cite{achiam2023gpt} or \claude \cite{claude3}, which exhibit enhanced abilities to comprehend content within applications compared to traditional models. 

A critical challenge we encountered was the absence of a suitable dataset for evaluating E2E test cases. To address this, we create a novel benchmark, called \benchname comprising \subjectcount open-source web applications. For each application, we extract all available features and employ instrumentation techniques to monitor front-end code, enabling us to track the actions performed on the web application during E2E tests. We establish a mapping between each application's features and the corresponding sequences of actions performed. As part of the benchmark, we develop a tool capable of automatically monitoring E2E test suite execution and assessing its coverage across all existing features within each application.

In summary, this work makes the following contributions:

\begin{itemize}[leftmargin=*]
    \item A formal formulation of feature-driven E2E test case generation, providing a theoretical foundation for developing practical, automated E2E testing tools.

    \item A probabilistic method for automatically inferring features in a web application. Our feature inference system assesses the likelihood of feature existence based on the frequency of the observed user actions.
    
    \item A novel technique, called \toolname, capable of autonomously generating feature-driven E2E test cases. Each test contains a sequence of actions to cover an inferred feature.
    
    \item \benchname, a novel benchmark designed for automatic evaluation of E2E test suites, quantifying their effectiveness in feature coverage.
\end{itemize}

Our evaluation results demonstrate that \toolname achieves a feature coverage of \ourscoverage, surpassing the best baseline, \crawljax, by \oursimprovement, and a significant \oursimprovementagent improvement over the best LLM agent-based baseline, \browsergym, highlighting the effectiveness of our methodology for automated E2E test generation. In addition to achieving superior coverage, \toolname generates more complex test cases. Furthermore, the test cases ranked as more likely to exist by our likelihood estimation system exhibit a higher correspondence with actual features within the web applications, further validating the effectiveness of our approach.
\section{Feature-Driven E2E Testing}
\label{sec:motivation}
To illustrate the challenges and opportunities in E2E test generation, we utilize the Amazon web application~\cite{amazon} as a motivating example, with a few representative pages illustrated in \autoref{fig:amazon-pages}. Before deploying such apps, developers must conduct rigorous testing to guarantee proper functionality. Testing can occur at various levels, including unit tests, integration tests, and E2E tests. E2E tests, in particular, focus on executing specific features and functionalities that users will engage with from start to finish, ensuring that different features and scenarios within the app behave as expected. Within the context of E2E testing, we define:

\begin{definition}[User Operation]
\label{def:user-operation}
A user operation, denoted by \(U\), is a sequence of user actions \(\{A_i\}\) (e.g., clicking, submitting forms). Each user operation can be classified into one of the following types:

\begin{itemize}[leftmargin=*]
    \item \textbf{Entity Operations:} These operations involve creating, reading, updating, or deleting data entities within the system. An entity operation is represented by a tuple \((X, E, M, P)\), where \(X\) signifies the CRUD action (Create, Read, Update, or Delete), \(E\) denotes the target entity or entity set, \(M\) is a boolean indicating whether the operation affects single or multiple entities, and \(P\) is a set of parameter values providing additional context or control.
    
    \item \textbf{Configuration Operations:} These operations modify system configurations or settings. They are represented by a tuple \((C, P)\), where \(C\) identifies the specific configuration being altered, and \(P\) specifies the new value.
\end{itemize}
\end{definition}

On the Amazon web application, searching for a product is classified as an entity operation, specifically a read operation \((X = \text{Read})\) on the Product entity set \((E = \text{Product})\). As a search typically returns multiple results, the multiplicity is true \((M = \text{True})\), and the parameter set includes the search term entered by the user \((P = \text{``search term"})\). It is important to distinguish between viewing a list of search results \((M = \text{True})\) and viewing the details of a single product \((M = \text{False})\), as these are distinct operations within the system.

Configuration operations are prevalent in various software systems. For example, logging into a system involves updating the \texttt{Authentication} configuration to reflect the user's authorized status. Toggling a \texttt{Dark Mode} setting modifies the application's visual appearance, while changing a \texttt{Language} configuration alters the language in which content is displayed to the user.

\begin{definition}[Application Feature]
\label{def:application-feature}
An application feature, denoted by $\mathcal{F}$, is characterized by the following properties:
\begin{itemize}[leftmargin=*]

    \item $\mathcal{F}$ can be realized through a finite ordered sequence of user operations (Definition \ref{def:user-operation}), denoted as $U_1 \rightarrow U_2 \rightarrow \cdots \rightarrow U_n$, where each $U_i$ represents a distinct user operation.
    
    \item Each user operation $U_i$ within the sequence is essential for achieving $\mathcal{F}$. Removing any $U_i$ would result in an altered or unattainable outcome.
    
    \item The outcome $O$ of executing the sequence $U_1, U_2, \dots, U_n$ is visually presented to the user upon completion, signifying the successful realization of $\mathcal{F}$.

    \item The label $L$ is an abstract, natural language description of the feature $\mathcal{F}$, independent of the parameters $P_i$ in the operations $U_i$.
\end{itemize}
\end{definition}

For instance, consider the common feature $\mathcal{F}$ on the Amazon web application labeled as \textit{``Adding a Product to the Shopping Cart"}. As illustrated in \autoref{fig:amazon-pages}, this feature entails a specific sequence of user operations $U_i$: (1) viewing (\textit{Read}) a list of products, (2) viewing (\textit{Read}) the details of a specific product, (3) Creating a cart item. This operation chain culminates in a visible change to the user's cart, representing the outcome $O$.

Each operation in this sequence is crucial for the successful completion of $\mathcal{F}$. For instance, removing the search operation would prevent the user from finding the desired product, while omitting the product selection step would leave the system without a specific item to add to the cart. In contrast, operations not essential to achieving the feature, such as changing the application's language, or background color should not be included in the defining sequence.

The abstract nature of $\mathcal{F}$ labeling implies that it should not be tied to specific parameters from the constituent operations. Whether the user searches for ``shoes" or ``electronics", or chooses a particular brand or model, the fundamental feature of \enquote{Adding a Product to the Shopping Cart} remains unchanged.

\header{Manual Testing}
In practice, manual E2E test creation often relies on developers or quality assurance (QA) engineers exploring the application or utilizing design documents to extract features, functionalities, and user flows within the application, outlining key scenarios to test. After specifying the features, test scripts for those features are often written using specialized frameworks like Selenium~\cite{selenium}, Cypress~\cite{cypress}, or Playwright~\cite{playwright} to simulate user interactions within the app. These scripts typically involve executing specific actions on the app, such as clicking buttons, filling out forms, and navigating through pages. They also include assertions to verify that the application responds as expected at each stage.

\header{Test Generation}
\label{sec:motivation-model}
Model-based testing has emerged as a prominent technique for automated E2E test generation in web applications~\cite{mesbah2012crawling, biagiola2019diversity, web-fragmentsRahul, web-matteo-icst20}. This approach involves systematically exploring the application's state space by exercising available actions, such as clicking links, hovering over elements, and submitting forms. The resulting state transitions are then captured and integrated into a model representing the application's behavior.

For instance, consider a model of the Amazon web application (illustrated in \autoref{fig:amazon-pages}). This model would encompass the states and action sequences necessary to add an item to the shopping cart. Furthermore, it would incorporate transitions facilitated by persistent navigation elements, such as the menu bar, enabling navigation between key pages like the landing page, login page, and shopping cart from any point within the application.
Once constructed, this model serves as the foundation for automatic test case generation. Test cases are derived as sequences of actions extracted from the model's transitions, with the objective of achieving comprehensive coverage based on predefined criteria, such as state coverage or transition coverage.


\autoref{lst:model-test-case} illustrates a model-based generated test case, showcasing a path within the Amazon web application.

\begin{lstlisting}[language=JavaScript, caption={Sample model-based generated test case}, label={lst:model-test-case}]
test("generated test case 1", () => {
    cy.visit("https://amazon.com");
    cy.get("search-bar").type("smartwatch");
    // assertion 1
    cy.get("login").click();
    // assertion 2
    cy.get("home").click();
    // assertion 3
    cy.get("cart").click();
    // assertion 4
});
\end{lstlisting}

This Cypress-based test simulates user interactions such as searching for a product (\enquote{smartwatch}), navigating from the search results to the login page, returning to the landing page, and finally proceeding to the cart page. At each stage, assertions are made to verify the correctness of the application's state, ensuring the presence of expected elements, validating displayed content, and confirming successful navigation.

\header{Challenges in E2E Testing}
Both manual test creation and automated test generation present distinct challenges. Developer-written test cases, while potentially comprehensive, are often expensive and time-consuming to develop. Model-based techniques offer a partial solution by automating test case generation. However, the resulting tests may lack the coherence and relevance of human-written scenarios.  This is evident in the sample model-based test case (\autoref{lst:model-test-case}), which, despite covering several states and transitions, the sequence of actions performed is not relevant to a specific application feature (Definition \ref{def:application-feature}). While developer-written tests typically follow meaningful features (e.g., adding an item to a cart), model-based tests prioritize coverage over coherence, leading to seemingly random sequences of actions. This trade-off between coverage and relevance highlights a key challenge in automated test generation. To more effectively evaluate the quality of generated E2E tests, we propose a new metric, feature coverage:

\begin{definition}[Feature Coverage]
\label{def:coverage}
Let $\mathcal{F} = \{f_1, f_2, \ldots, f_m\}$ be the set of all features (Definition \ref{def:application-feature}) in an application, and let $\mathcal{T} = \{t_1, t_2, \ldots, t_n\}$ be a set of test cases in a test suite designed to test these features. Define a relation $R \subseteq \mathcal{T} \times \mathcal{F}$ where each $(t_i, f_j) \in R$ indicates that test case $t_i$ exercises feature $f_j$. Feature coverage $C$ is then defined as the ratio of the number of unique features exercised by the test suite to the total number of features, given by:
\[
C = \frac{|\{ f_j \in \mathcal{F} \mid \exists t_i \in \mathcal{T} \text{ such that } (t_i, f_j) \in R \}|}{|\mathcal{F}|}
\]

\noindent
where each test case $t_i$ targets exactly one feature $f_j$ and it is permissible for multiple test cases to cover the same feature.
\end{definition}

This metric shifts the focus from purely syntactical (code coverage) or structural coverage (such as state and transition coverage) to a more user-centric perspective, emphasizing the testing of distinct application features. Automating test case creation while achieving a high feature coverage necessitates a nuanced understanding of the application's context and content, a capability that remains absent in existing automation techniques. To address this gap, a novel feature-driven E2E test generation approach is required. This approach must be capable of inferring the features existing within the app, connecting the available actions to those features, and generating the sequence of actions corresponding to the coverage of each feature as a test case.
\section{Approach}
\label{sec:approach}
\begin{figure*}[t]
    \centering
    \includegraphics[width=\textwidth]{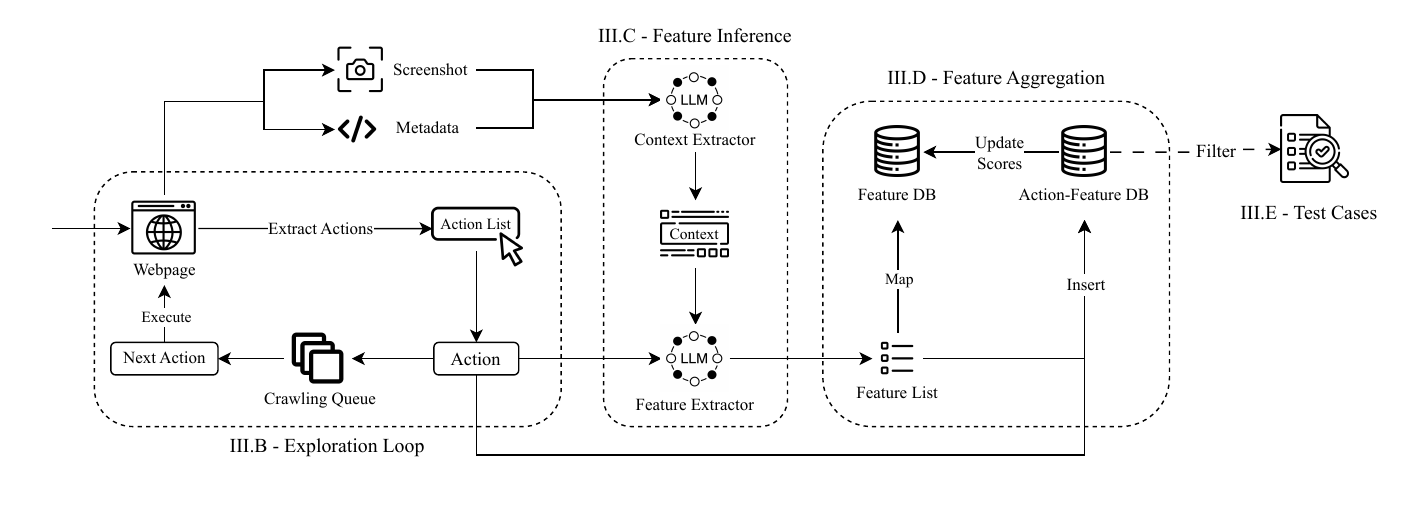}
    \caption{Overview of our framework}
    \label{fig:workflow}
\end{figure*}

In this work, we introduce \toolname, a novel approach for automatically generating semantically meaningful E2E tests, each targeting a distinct application feature (Definition \ref{def:application-feature}), aiming to achieve high feature coverage (Definition \ref{def:coverage}). Our primary focus is addressing the challenge of inferring application features from the contextual information present in various application states. We propose a method that assesses the likelihood of features existing within an application based on observed user actions. By integrating this method with LLMs, we establish a workflow to identify the features present in the application and the corresponding sequences of actions required to trigger them. These action sequences are then transformed into comprehensive test cases for the application. The architecture for \toolname is illustrated in \autoref{fig:workflow}.

\subsection{Feature Inference Modeling}
\label{sec:modeling}
We first formalize the feature inference task in order to leverage LLMs more effectively. A web app typically consists of various application states and transitions between them:

\begin{definition}[Application States and Transitions]
\label{def:application-state}
 An application state $S_i$ represents a snapshot of the web app at a particular moment, characterized by the runtime values of relevant variables, as well as the dynamic structure and content as rendered in the browser. A transition $A_i$ initiated by a user action (e.g., clicking a button, submitting a form) can cause a state change, e.g., from $S_1$ to $S_2$.
\end{definition}

\autoref{fig:amazon-pages} provides a visual representation of this concept, where each image depicts a distinct state within the Amazon web application.  Furthermore, the actions available within these images represent potential transitions to other states.

Given a web app with $K$ states, $\{S_1, S_2, \ldots, S_K\}$, and $M$ potential features, $\{\mathcal{F}_1, \mathcal{F}_2, \ldots, \mathcal{F}_M\}$, the task of inferring features becomes that of constructing a generative model to estimate the following distribution:
\begin{equation}
    \label{eq:initial-model}
    p(\mathcal{F}_1, \mathcal{F}_2, \ldots, \mathcal{F}_M | S_1, S_2, \ldots, S_K) = p(\textbf{F} | \textbf{S})
\end{equation}

In this formulation, the probability distribution $p(\textbf{F} | \textbf{S})$ represents the likelihood of a feature set $\textbf{F}$ being present within an application, given the information observed in the set of states $\textbf{S}$. A generative model could learn this distribution from data and subsequently generate a feature set $\textbf{F}$ that maximizes this probability.

This approach aligns with how humans typically extract features and design test scenarios. Users explore the application's interface, inferring available features based on observations. For instance, when presented with the application states depicted in \autoref{fig:amazon-pages}, a human can intuitively infer the presence of features such as searching for products, purchasing products, viewing account details, and reviewing order details. These inferences are based on the visual cues and interactive elements present in the observed states, leading to a higher likelihood of these features being available within the Amazon web application.

However, constructing a generative model capable of accurately estimating $p(\textbf{F} | \textbf{S})$ is a challenging task due to the vast complexity of both the feature and state space. To address this challenge, we introduce a simplified model that enhances tractability while retaining the essence of the generative approach.

\subsubsection{Feature Independence}
Given complete access to all states \(\textbf{S}\) within an application, the inference of individual features \(\mathcal{F}_i\) can be considered independent. While certain features may often imply the existence of others (e.g., an \enquote{add to cart} feature suggests a \enquote{remove from cart} functionality), knowledge of the complete state space allows for direct observation and inference. For instance, the presence of a \enquote{remove} button on the Cart page confirms the existence of the \enquote{remove from cart} feature, independent of knowledge about the \enquote{add to cart} feature. Leveraging this feature independence given $\textbf{S}$, the joint probability distribution in \autoref{eq:initial-model} simplifies to:
\[
    p(\textbf{F} | \textbf{S}) = p(\mathcal{F}_1 | \textbf{S}) p(\mathcal{F}_2 | \textbf{S}) \ldots p(\mathcal{F}_M | \textbf{S})
\]

Based on this notion, instead of attempting to infer all features simultaneously, we can leverage the conditional independence and generate features individually using the distribution $p(\mathcal{F}_i | \textbf{S})$. By sampling the top $M$ generated values from this distribution, we can effectively identify the most probable existing features within the application.

\subsubsection{Action-Centric Feature Inference}
Feature determination in web applications can typically be accomplished through an action-centric lens, focusing on the available actions on a page rather than the specific content. As illustrated in Figure~\ref{fig:amazon-page-4}, the \enquote{Add to Cart} feature on a product page is discernible solely from the context of the page and the presence of the corresponding action, regardless of the product's details. Having a general context for the page is particularly valuable when dealing with actions that have ambiguous descriptions, such as a ``continue" button. In such cases, the context of the current page (e.g., checkout page) aids in clarifying the intended purpose of the action. By adopting this action-centric perspective, we shift the focus from analyzing the entirety of the application's state information to a more targeted examination of the actions and their associated contextual information. Formally, we can express this as:
\begin{equation}
    \label{eq:action-centric-model}
    p(\mathcal{F}|\textbf{S}) = p(\mathcal{F} | A_{1,1}, A_{1,2}, \ldots, A_{K, n_K})
\end{equation}
where $A_{i, j}$ and $n_i$ represent the $j$-th action and the number of actions on state $S_i$ respectively.

\subsubsection{Sequential Action Chains}
The features within web apps are designed to be executed through a sequential chain of actions, as can be observed in Definitions \ref{def:user-operation} and \ref{def:application-feature}. For a given feature $\mathcal{F}$, there exists an ordered sequence of actions ${A_1, A_2, \ldots, A_N}$ that leads to its execution: 
\[
    \mathcal{F}: A_1 \rightarrow A_2 \rightarrow \ldots \rightarrow A_N
\]

Crucially, a feature should be derivable solely from its action chain. For example, when adding an item to a shopping cart, the presence of actions for rating or commenting on products is irrelevant.  This allows us to eliminate unrelated actions from the context of \autoref{eq:action-centric-model}, focusing solely on the actions $\{A_i\}$ relevant to the feature:
\[
    p(F | \textbf{S}) = p(F | A_1, A_2, \ldots, A_N)
\]
Then we can use Bayes' theorem and chain rule of probabilities to simplify this to:
\[
    = \frac{p(F, A_1, A_2, \ldots, A_N)}{p(A_1, A_2, \ldots, A_N)}
\]
\begin{equation}
    \label{eq:bayes-simplified}
    = \frac{p(F)p(A_1 | F)p(A_2 | A_1, F)\ldots p(A_N | A_1, \ldots, F)}{p(A_1, A_2, \ldots, A_N)}
\end{equation}

\subsubsection{Dependence of Subsequent Actions}
The sequential nature of user interactions within web applications can be modeled by recognizing the dual dependency of each action: an action $A_i$ depends on the immediately preceding action $A_{i - 1}$ and the specific feature $\mathcal{F}$ the user intends to execute. This dependency allows us to estimate the most probable subsequent actions based on the current action and feature.

Consider the scenario illustrated in \autoref{fig:amazon-page-3}, where a user clicks on a product within a list of search results. If the user intends to purchase the product (feature $\mathcal{F}$), we can predict that the next likely action would be clicking on \enquote{Add to Cart.} Conversely, if the user intends to rate the product (a different feature $\mathcal{F}$), the most probable next action would be clicking on a rating value. Crucially, this estimation of the next action relies solely on the current action and the intended feature, regardless of the user's prior interactions and the specific path taken to reach the current state. Whether the user arrived at the product page through searching, browsing categories, or any other means is irrelevant. Consequently, based on this property, we can simplify \autoref{eq:bayes-simplified}:
\[
    p(\mathcal{F} | \textbf{S}) = \alpha(\textbf{A}) p(\mathcal{F}) p(A_1 | \mathcal{F}) \prod_{i=2}^{N} p(A_i | A_{i - 1}, \mathcal{F})
\]
\[
    p(\mathcal{F}) p(A_1 | \mathcal{F}) = p(A_1) p(\mathcal{F} | A_1)
\]
\[
    p(A_i | A_{i - 1}, \mathcal{F}) = \frac{p(\mathcal{F} | A_i, A_{i - 1}) p(A_i | A_{i - 1})}{p(\mathcal{F} | A_{i - 1})}
\]
\begin{equation}
    \rightarrow p(\mathcal{F} | \textbf{S}) = \beta(\textbf{A}) p(\mathcal{F} | A_1) \prod_{i = 2}^{N} \frac{p(\mathcal{F} | A_i, A_{i - 1})}{p(F | A_{i - 1})}
\end{equation}

In the presented equations, $\alpha(\textbf{A})$ and $\beta(\textbf{A})$ denote generalized action probabilities. These terms are not specific to any particular application or feature but rather capture the inherent likelihood of observing certain actions across all the different web interactions. For instance, these terms might encapsulate the probability of encountering a ``Login" action in any web application, essentially providing a baseline expectation for the occurrence of actions. Since our goal is to maximize the distribution over $F$, these functions become irrelevant. Therefore, we can further simplify the task of feature inference into the following equation:
\begin{align}
    \label{eq:final-model}
    \mathcal{F} &= \arg\max_\mathcal{F} \sum_{i=1}^{N} \Big( \log\big(p(\mathcal{F} \mid A_i, A_{i - 1})\big) \notag\\
    &\quad - \log\big(p(\mathcal{F} \mid A_{i - 1})\big) \Big)
\end{align}

This result has an intuitive interpretation. If we are predicting the existence of a certain feature based on an action, we should observe further evidence supporting that feature after performing the action.

\autoref{eq:final-model} provides a flexible foundation for various implementations, as it applies broadly to web applications and is not constrained by any specific implementation details in its derivation. In this work, we implement our method based on this equation by utilizing LLMs to estimate the distributions $p(F | A_i)$ and $p(F | A_i, A_{i - 1})$. This is achieved by feeding $\{A_i\}$ and $\{A_{i - 1}, A_{i}\}$ as context to the LLM, respectively, and prompting it to infer features based on this context. We then aggregate the generated results to assess the likelihood of existing features within the application. The remaining sections of our approach will detail how we leverage LLMs and aggregate their outputs to infer both features and their corresponding chains of actions.

\subsection{Exploration Loop}
As illustrated in \autoref{fig:workflow}, \toolname operates on an exploration loop paradigm, interfacing with the target web application, capturing user actions, and systematically queuing them for execution. Each executed action potentially reveals new application states, driving exploration until no further novel states are discovered or a timeout occurs.
\toolname employs a breadth-first search (BFS) strategy for state exploration, queuing, and recursively crawling neighboring states from the current state. Feature extraction inferences are performed concurrently during state visits, feeding the extracted actions into the rest of the workflow to infer features.

\subsection{Feature Inference}
\label{sec:feature-inference}
As the exploration loop discovers new states and extracts their associated actions, \toolname concurrently analyzes each action to infer the specific features with which it interacts. For the following sections, imagine our exploration has led us to state $S_i$ via the action sequence $A_1 \rightarrow \ldots \rightarrow A_{i-1}$. In this state, we observe a set of available actions $\textbf{A}_i = \{A_{i1}, A_{i2}, \ldots, A_{in}\}$, where $A_{ij}$ denotes the $j$th action on $S_i$.

\subsubsection{State Context Extraction}
As discussed in \autoref{eq:action-centric-model}, actions are represented alongside their corresponding context—the high-level purpose of the page where the action occurs. This contextual information is essential for accurate feature inference, particularly when the content associated with an action is ambiguous (e.g., a button labeled ``Continue") and requires further disambiguation.

Identifying page context requires multiple data sources. The application's description and category provide high-level information, while the page's content (text, HTML, images) offers more specific details. We prioritize image-based analysis over HTML due to HTML's verbosity. Additionally, the history of actions leading to the current page provides further contextual clues. Consequently, our context extraction employs a multi-modal approach, incorporating a screenshot of the $S_i$ rendered in a browser, a description of the entire application, and the most immediate action $A_{i - 1}$ leading to $S_i$. As an example, the following is the extracted context for the page in \autoref{fig:amazon-page-3}: \textit{A webpage displaying search results for a product query, allowing users to browse and filter options for purchasing.}

\subsubsection{Feature Extraction}
Following the extraction of contextual information from $S_i$, we can now utilize that context in conjunction with the actions present on the state, $\textbf{A}_i$, and prompt an LLM to predict possible features connected to each action. We follow the result in \autoref{eq:final-model} derived in Section \ref{sec:modeling} to infer the features by querying the LLM twice. The first prompt requests the LLM to generate features based on the individual actions in $S_i$, corresponding to \(p(\mathcal{F} | A_{ij} \in \textbf{A}_i)\). The second prompt asks the LLM to generate features based on the actions in $S_i$ plus the most recent action that led to the current state, representing \(p(\mathcal{F} | A_{ij} \in \textbf{A}_i, A_{i - 1})\). To illustrate this process, consider \autoref{fig:amazon-page-4}.  For the first prompt, the LLM would be asked to generate features solely based on the ``Add to Cart" button.  However, for the second prompt, the LLM would consider both the ``Add to Cart" button and the preceding action that led to this page, which is clicking the product link in \autoref{fig:amazon-page-3}. 

Estimating the probability of a result generated by an LLM, particularly in proprietary models, is not always feasible. This limitation hinders the direct calculation of \(p(\mathcal{F} | A_{ij} \in \textbf{A}_i)\) and \(p(\mathcal{F} | A_{ij} \in \textbf{A}_i, A_{i - 1})\) in \autoref{eq:final-model}. To address this, we incorporate Chain of Thought (CoT) prompting in the LLM prompt, asking it to generate a list of \(R\) features ordered by their perceived probability of existence. CoT prompting encourages the LLM to provide more reasoned and reliable responses, increasing the validity of the feature ordering. Having access to this ordering, we employ a geometric distribution to estimate the probability of an inferred feature based on its rank:
\[
    \text{rank\_score}(r | r \leq R) = \log\big(p(\mathcal{F} \text{ is ranked } r)\big)
\]
\begin{equation}
    \label{eq:prob-rank}
     = \log\big((1 - p)^{r- 1} p\big) = (r - 1) \log(1 - p) + log(p)
\end{equation}
where \(p\) is a manually set parameter. A $p$ value close to 1 creates a significant difference in probability between the top-ranked item and the rest, while a $p$ value near 0 assigns nearly equal probability to each rank. As there is often more than one feature associated with an action, we seek a balance that allows our model to recognize these multiple features while still distinguishing between higher and lower-ranked items. Therefore, we set \(p\) to 0.5.

Since the list of our inferred features is limited to the top $R$, we must also take into account the probability that a feature is not in the top $R$, but appears at a certain rank if we extend our list. For every feature that does not appear in the top $R$ inferred features, we use a constant score:
\begin{equation}
    \label{eq:constant-prob-rank}
     \text{rank\_score}(r | r > R) = R \log(1 - p) + log(p)
\end{equation}

This constant score is used in the aggregation phase.

\subsection{Feature Aggregation}
During the exploration process, upon encountering a new state, we extract potential features associated with each action within that state. Importantly, these feature inferences are conducted in isolation. Features derived using the current action context $\{A_{ij}\}$ are independent of those inferred using both the current and preceding actions $\{A_{ij}, A_{i-1}\}$. Furthermore, both sets of newly inferred features are initially disconnected from any previously generated features. To reconcile these disparate inferences, we enter the aggregation phase. To manage the aggregation, we employ two interconnected databases.

\subsubsection{Feature Database (\funcdb)}
The Feature Database (\funcdb) is a vector database storing a global list of discovered features. Each entry in \funcdb contains a label, i.e., a textual description of the feature, its corresponding embedding, and a confidence score derived from Equation~\ref{eq:final-model}. This score reflects the likelihood of the feature's existence within the application.

\subsubsection{Action-Feature Database (\actionfuncdb)}
\funcdb is complemented by the Action-Feature Database (\actionfuncdb). Each row in \actionfuncdb associates an action within a state with its corresponding inferred features from \funcdb. Additionally, it records a rank score calculated using Equation~\ref{eq:prob-rank} and indicates whether the inference utilized solely the current action \(A_{ij}\) or both the current state's action and preceding action \(A_{ij}, A_{i-1}\) as context.

The information stored in \funcdb and \actionfuncdb is dynamic, evolving as the application is explored. These databases continuously interact, influencing each other to update the overall confidence scores of features. This iterative refinement ensures that the feature model remains accurate and comprehensive as new observations are made.

\subsubsection{Mapping inferred features to \funcdb}
To update feature scores, we first map the newly inferred features to existing entries in the \funcdb. This is achieved by leveraging the textual embeddings of feature descriptions stored in \funcdb. We query \funcdb using the embedding of an inferred feature and retrieve the top results based on cosine similarity. This metric identifies features in \funcdb that are semantically closest to the inferred feature.
By establishing these connections, we can then update the confidence scores of the corresponding features in \funcdb, incorporating the information gained from the new observations.

While querying the \funcdb using textual embeddings can identify semantically similar features, it does not guarantee a perfect match.  To ensure accuracy, we introduce an additional validation step using an LLM.  This step involves querying the LLM to assess whether the inferred feature aligns with any of the similar feature descriptions retrieved from \funcdb. The LLM acts as a semantic arbiter, determining if any of the retrieved descriptions truly describe the same feature.  \autoref{lst:mapping-query} provides a concrete example of this process, showcasing how the LLM helps refine the initial matches obtained from \funcdb for an ``Add to Cart" action's feature in \autoref{fig:amazon-page-4}. 

\begin{lstlisting}[language=JavaScript, caption={Example for \funcdb querying}, label={lst:mapping-query}]
// A feature inferred for an action
"Add a product to the cart"

// FD retrieval results
"Adding to the cart"          // exact match
"View product details"        // mismatch
"Remove item from the cart"   // mismatch
"Explore product categories"  // mismatch
"Navigate to homepage"        // mismatch
\end{lstlisting}

If no match is found in \funcdb, this indicates that the feature is a novel observation within the application. In such cases, the feature data, along with an initial confidence score of 0, is inserted into \funcdb.  Following the matching (or insertion) of a feature in \funcdb, we then create a corresponding entry in the \actionfuncdb. This entry includes a pointer to the relevant \funcdb entry, establishing the connection between the two databases. With these updates in place, we can then proceed to refine the confidence score of the feature in \funcdb, incorporating the information gained from the new observation.

\subsubsection{Updating the Scores}
For each action $A_{ij}$, we have inferred a corresponding set of features $\{\mathcal{F}_{j1}, \mathcal{F}_{j2}, \ldots\}$, as detailed in Section \ref{sec:feature-inference}. To update the feature scores based on these new observations, we employ the following formula, derived from Equations~\ref{eq:final-model} and \ref{eq:prob-rank}:
\[
  \text{score\_update}(F_{jk}) = \text{rank\_score}(F_{jk} | A_{ij}, A_{i - 1}) -
\]
\begin{equation}
  \label{eq:score-incremenet}
  \text{rank\_score}(F_{jk} | A_{i - 1})
\end{equation}

This formula calculates the change in the score for feature $F_{jk}$ (the $k$th feature associated with action $A_{ij}$) based on the observed action sequence. The first term, $\text{rank\_score}(F_{jk} | A_{ij}, A_{i - 1})$, represents the probability of the feature given both the current and preceding actions. The second term, $\text{rank\_score}(F_{jk} | A_{i - 1})$, represents the probability of the feature given only the preceding action. The difference between these two terms reveals the incremental impact of action $A_{ij}$ on our confidence in the existence of feature $F_{jk}$.

To address cases where feature $F_{jk}$ exists in the LLM's inference for ($A_{ij}$, $A_{i-1}$) but not for $A_{i-1}$, we substitute the $\text{rank\_score}(F_{jk} | A_{i - 1})$ term in Equation~\ref{eq:score-incremenet} with the constant ranking score from Equation~\ref{eq:constant-prob-rank}. Using the $score\_update$ formula, we then locate $F_{jk}$'s corresponding entry in \funcdb and update its score. This process is repeated throughout the app exploration.

\subsection{Generating Test Cases}
\label{sec:gen-test-case}
Following the exploration phase, we sort the features within the \funcdb. From this sorted list, we filter and retain the top-scoring features as those identified within the application. This filtering process involves selecting a lower bound cutoff, determined by \autoref{eq:prob-rank} as $\log(p)$. This choice of cutoff is motivated by the possibility of single-action features, which could be correctly predicted by the LLM with a rank of 1 and a corresponding score of $\log(p)$.  By setting the cutoff at this value, we ensure the inclusion of such features in our generated tests.

After filtering, we then extract the corresponding chains of actions from the \actionfuncdb and translate them into test cases. This process leverages the accumulated evidence gathered during exploration to identify the most likely features and their associated action sequences, ultimately generating test cases that semantically cover the application features.


\subsection{Implementation}
\toolname is implemented in Python using the LangChain framework \cite{langchain}, offering flexibility in LLM selection. For our evaluations, we opted for \sonnet due to its recognized performance among the most advanced LLMs. Textual embeddings for feature descriptions and \funcdb utilize the ADA architecture \cite{neelakantan2022text}, and the E2E tests are generated in Selenium~\cite{selenium}. Our databases, \funcdb and \actionfuncdb, are hosted on MongoDB Atlas\footnote{\href{https://www.mongodb.com/}{https://www.mongodb.com/}}, which provides a vector search functionality well-suited for querying the feature description embeddings.
\section{Benchmark Construction}
\label{sec:benchmark}
Assessing E2E test cases presents a significant challenge in software testing automation. To the best of our knowledge, there is an absence of datasets for E2E test case evaluation that hinders further progress in this area of study. To address this gap, we embarked on the creation of such a benchmark, called \benchname. This section elucidates the complexities inherent in dataset creation and outlines the methodological steps undertaken to achieve this goal.

\header{Benchmark Construction Challenges}
The lack of a benchmark for assessing Feature Coverage in E2E test cases arises from the inherent challenges of evaluating this metric. As evident in Definition \ref{def:coverage}, this evaluation hinges on two critical steps: identifying the features within an application and mapping E2E test cases to these features to measure coverage.

The first challenge in this process is the identification and quantification of features within a web application. Features are often difficult to define and vary widely depending on the context, making their extraction difficult and subjective. Once features are identified, the second challenge is determining which specific feature a test case targets. This becomes particularly complex in large-scale applications like Amazon (illustrated in \autoref{fig:amazon-pages}), where a feature such as \enquote{Viewing the product's details} can be accessed through multiple pathways—whether by clicking on a product from the landing page, filtering through categories, or using the search function. As the number of available actions increases, so does the complexity of tracking these diverse paths. For example, the Amazon web application with millions of products could have numerous valid E2E test cases for viewing product details, each following a distinct path. An effective evaluation tool must be capable of accounting for and managing this inherent path redundancy.

\header{Feature Identification}
Regarding the first challenge, we have established a precise and formal definition of a feature (Definition \ref{def:application-feature}) in this work, providing a foundation for systematic and objective identification of features within software applications.

\header{Instrumentation for Feature Mapping}
To address the second challenge in mapping test cases to features, we need to track user (i.e., E2E test) interactions within web applications. To automate this tracking, we select open-source web applications for our benchmark, enabling us to directly instrument their code. The instrumentation adds logging mechanisms to capture every user action within the app, encompassing clicks, hovers, and various input types (selects, texts, radio buttons, checkboxes). Each unique action component within the app generates a distinct log message containing a unique identifier. 
To handle path redundancy, we leverage the inherent modularity of web applications.
For example, consider an open-source e-commerce application: products are typically rendered using a collection of \code{Product} components. By leveraging this component-based structure, we can instrument component code to track the interactions by the E2E test cases. This approach effectively circumvents the challenge of near-infinite pathways, as each action within a path is implemented within a discrete component, regardless of its frequency of occurrence.

\header{Feature Grammar Extraction}
With the tracking system established, we then map the chain of logs to specific features within each subject application. This process involves identifying the available features for each subject in the benchmark and executing sequences of actions to trigger them while tracking the corresponding logs. These logs are subsequently transformed into a \textit{feature grammar}, which serves as a ground-truth reference for assessing E2E test cases. To illustrate, a sample feature grammar for purchasing a product in an e-commerce application would be the following:

\begin{lstlisting}[language=JavaScript, caption={Feature Grammar for ``product purchase''}, label={lst:saleor-grammar}]
"c1-product-element" "c10-add-to-cart"
"c22-cart-icon" "c12-checkout-button"
("t5-credit-number" | "t16-credit-date" | "t23-credit-cvv")+ "c35-complete-purchase"
\end{lstlisting}

In this feature grammar, the sequence of actions begins by clicking on the product, followed by adding it to the cart and navigating to the cart page. Subsequently, the checkout button is clicked, leading to the purchase page where credit card information is required. The input fields for credit card details can be filled in any order, represented by the logical OR operator. Additionally, the information in the text boxes can be modified multiple times, denoted by the ``+" sign indicating repetition. Finally, the purchase is completed by clicking a confirmation button.

The extraction of feature grammar for each subject was conducted independently by each author, followed by a collaborative discussion to consolidate the findings. This iterative process ensured a comprehensive identification of features, with consensus reached on both the features themselves and any alternative paths to achieve the same functionality. The resulting grammars then serve as a basis for evaluating the coverage of a test suite over all of the functionalities.

\header{Benchmark Subjects}
A list of the apps in our benchmark is available in \autoref{tab:dataset}. These applications, most of which have been previously used in web testing research \cite{yandrapally2022fragment, yandrapally2023carving, stocco2016clustering, biagiola2019web}, encompass a diverse range of categories, including bug tracking (\textit{MantisBT}), e-commerce (\textit{Saleor}), and translation management (\textit{EverTraduora}).

\header{Automatic Coverage Evaluation}
During test case execution, our benchmark continuously monitors the performed actions, trying to map the sequence to one of the identified functionalities within the application. This allows us to assess the coverage of an E2E test suite across all the distinct features in our benchmark subjects. The calculation of feature coverage is performed completely automatically based on the grammar extracted in the previous phase, resulting in an objective measurement of the Feature Coverage (Definition \autoref{def:coverage}).

\section{Evaluation}
\label{sec:evaluation}

We have framed the following research questions to measure the effectiveness of \toolname:

\begin{itemize}[leftmargin=*]
\item \textbf{RQ1}: How effective is \toolname in generating feature-driven E2E tests?

\item \textbf{RQ2}: How accurate is the feature inference of \toolname?

\item \textbf{RQ3}: How does \toolname compare to other state-of-the-art techniques?
\end{itemize}

\header{Process}
We use \benchname to evaluate the efficacy of \toolname and other methods in achieving feature coverage. For running our experiments, we set the temperature parameter of the LLMs to 0 to produce the same response every time. 

\begin{table}[t]
{\small
    \caption{Benchmark Subjects}
    \label{tab:dataset}
    \centering
    \begin{tabular}{llrr}
        \toprule
        \textbf{App Name} & \textbf{Category} & \textbf{Features} & \textbf{LOC} \\
        \midrule

        \rowcolor{lightgray}
        PetClinic & Health & 23 & 51K \\
        
        Conduit & Blog & 17 & 53K \\

        \rowcolor{lightgray}
        Taskcafe & Task Manager & 32 & 67K \\
        
        Dimeshift & Expense Tracker & 21 & 10K \\

        \rowcolor{lightgray}
        MantisBT & Bug Tracker & 27 & 118K \\

        EverTraduora & Translation Manager & 41 & 25K \\

        \rowcolor{lightgray}
        Saleor Storefront & E-Commerce & 13 & 58K \\

        Saleor Dashboard & E-Commerce Admin & 130 & 1.1M \\
        \bottomrule
    \end{tabular}
}
\end{table}

\header{Baselines}
To the best of our knowledge, no existing technique directly addresses feature-driven E2E test generation. Recent advances in LLM-based agents such as \webcanvas \cite{pan:mind2web-live:arxiv2024} and \browsergym \cite{workarena2024} have demonstrated the potential to navigate web applications and execute user instructions. However, adapting these agents for E2E test generation poses several challenges. First, they often require pre-existing feature extraction, which is the primary focus of our approach. Second, these agents may struggle to interpret abstract task descriptions and instead require concrete, detailed instructions for execution.

To assess the effectiveness of our proposed methodology, we evaluate \toolname against these specific-purpose agents, \webcanvas and \browsergym, both of which utilize \gpt as their underlying LLM. We also compare \toolname with more generalized agents: AutoGPT \cite{autogpt2024}, which employs \gpt for both task planning and execution, and OpenDevin \cite{opendevin2024}, a code-focused agent that utilizes the \claude LLM. Additionally, we benchmark against a model-based technique, represented by \crawljax's test generation module \cite{mesbah2012crawling, web-fragmentsRahul}. All agents are instructed to navigate web applications and generate end-to-end (E2E) test cases.

\subsection{Effectiveness (RQ1)}
The primary objective of our method is to maximize the extent of \textit{Feature Coverage} as defined in Definition \ref{def:coverage}. As described in Section \ref{sec:approach}, \toolname generates a set of E2E test cases. We evaluate the generated tests using the methodology detailed in \autoref{sec:benchmark}.

\toolname demonstrates the ability to generate test cases that cover an average of \ourscoverage of features across the applications in the \benchname benchmark. Considering all features across all applications, \toolname achieves a total Feature Coverage of \totalfeaturecoverage. A more granular breakdown of the feature coverage for each app is presented in the \textit{Recall} column of \autoref{tab:inference-stats}. These results underscore the effectiveness of \toolname in effectively inferring features and generating corresponding test cases across a diverse range of applications. 

\subsection{Feature Inference (RQ2)}
As described in Section \ref{sec:approach}, \toolname infers a list of features along with their scores by the end of the inference phase.
\autoref{tab:inference-stats} provides statistics for both correct and incorrect predictions within the inferred feature list.
\toolname achieves a total precision of 0.55, indicating that 55\% of the generated test cases were correct based on the evaluation metrics. The total recall of the model, which is the same as the average Feature Coverage across all features in all applications, reaches 0.72. This means that our generated test cases successfully cover \totalfeaturecoverage of the total features present. The precision and recall values result in an F1 score of 0.62, demonstrating a reasonable balance between \toolname's ability to identify relevant test cases and its ability to capture the full spectrum of application features. 

\begin{table}[h]
{\small
    \caption{Inference Statistics}
    \label{tab:inference-stats}
    \centering
    \begin{tabular}{lrrrrr}
        \toprule
        \textbf{App Name} & \textbf{Total} & \textbf{Correct} & \textbf{Precision} & \textbf{Recall} & \textbf{F1} \\
        \midrule

        \rowcolor{lightgray}
        PetClinic & 29 & 19 & 0.66 & 0.83 & 0.73 \\

        Conduit & 19 & 14 & 0.67 & 0.82 & 0.76 \\
        
        \rowcolor{lightgray}
        Taskcafe & 50 & 22 & 0.42 & 0.66 & 0.51 \\

        Dimeshift & 40 & 17 & 0.42 & 0.81 & 0.56 \\

        \rowcolor{lightgray}
        MantisBT & 45 & 22 & 0.47 & 0.75 & 0.58 \\

        EverTraduora & 69 & 30 & 0.42 & 0.73 & 0.54 \\

        \rowcolor{lightgray}
        Storefront & 21 & 13 & 0.62 & 1.0 & 0.76 \\

        Dashboard & 123 & 85 & 0.69 & 0.65 & 0.67 \\

        \midrule
        Total & 396 & 222 & 0.55 & 0.72 & 0.62 \\
        
        \bottomrule
    \end{tabular}
}
\end{table}

However, the effectiveness of our approach is not solely determined by the number of correct features. As detailed in Section \ref{sec:approach}, we employ a scoring and filtering system to prioritize feature generation and test case creation.  If this system functions as intended, features with higher scores should be more likely to correspond to actual features within the application.

\begin{figure}[h]
    \centering
    \includegraphics[width=0.48\textwidth]{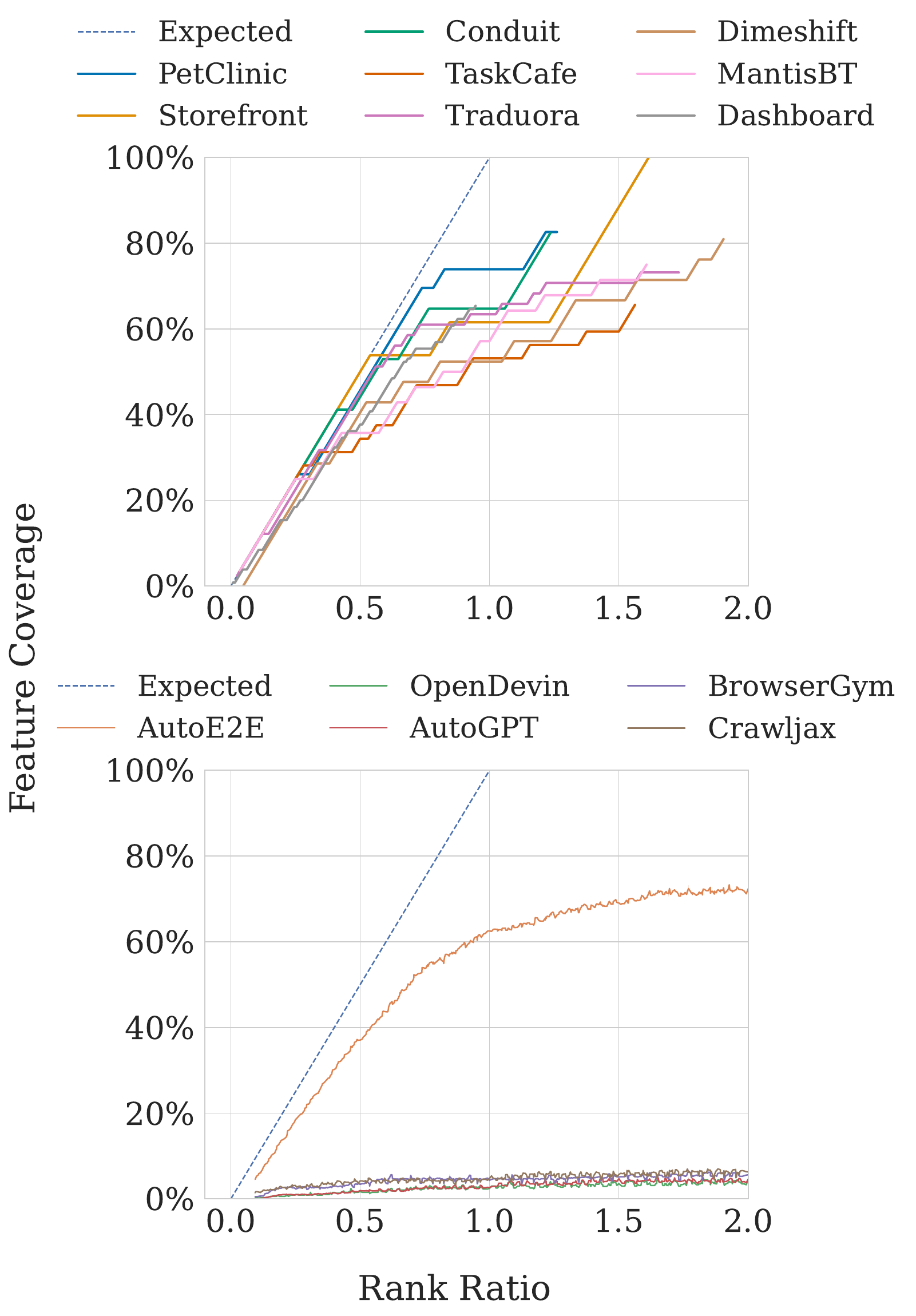}
    \caption{Feature Coverage vs. rank ratio}
    \label{fig:rank-coverage}
\end{figure}

\autoref{fig:rank-coverage} illustrates the coverage of top-$k$ features in relation to the ratio of $k$ to the total number of features present in the application, referred to as the \textit{Rank Ratio}. The left half of this figure demonstrates the coverage against the rank ratio for \toolname, while the lower half displays the moving average of coverage for \toolname and the baseline methods. In an ideal scenario, where all top-$k$ inferred features are accurately identified, the plotted line would follow a 45-degree trajectory. The figure reveals the extent of coverage achieved as the feature list expands to encompass more candidates.

As evident in the figure, \toolname generally follows the expected 45-degree line up to a rank ratio of approximately 0.75. This indicates that, on average, for an application with \(N\) features, the first \(0.75N\) generated test cases are almost all correct. However, a divergence is observed beyond this point, suggesting that while the remaining generated test cases may be correct in certain instances, they do not consistently cover features with the same level of accuracy.

\subsection{Comparison (RQ3)}
\begin{figure}
    \centering
    \includegraphics[width=0.48\textwidth]{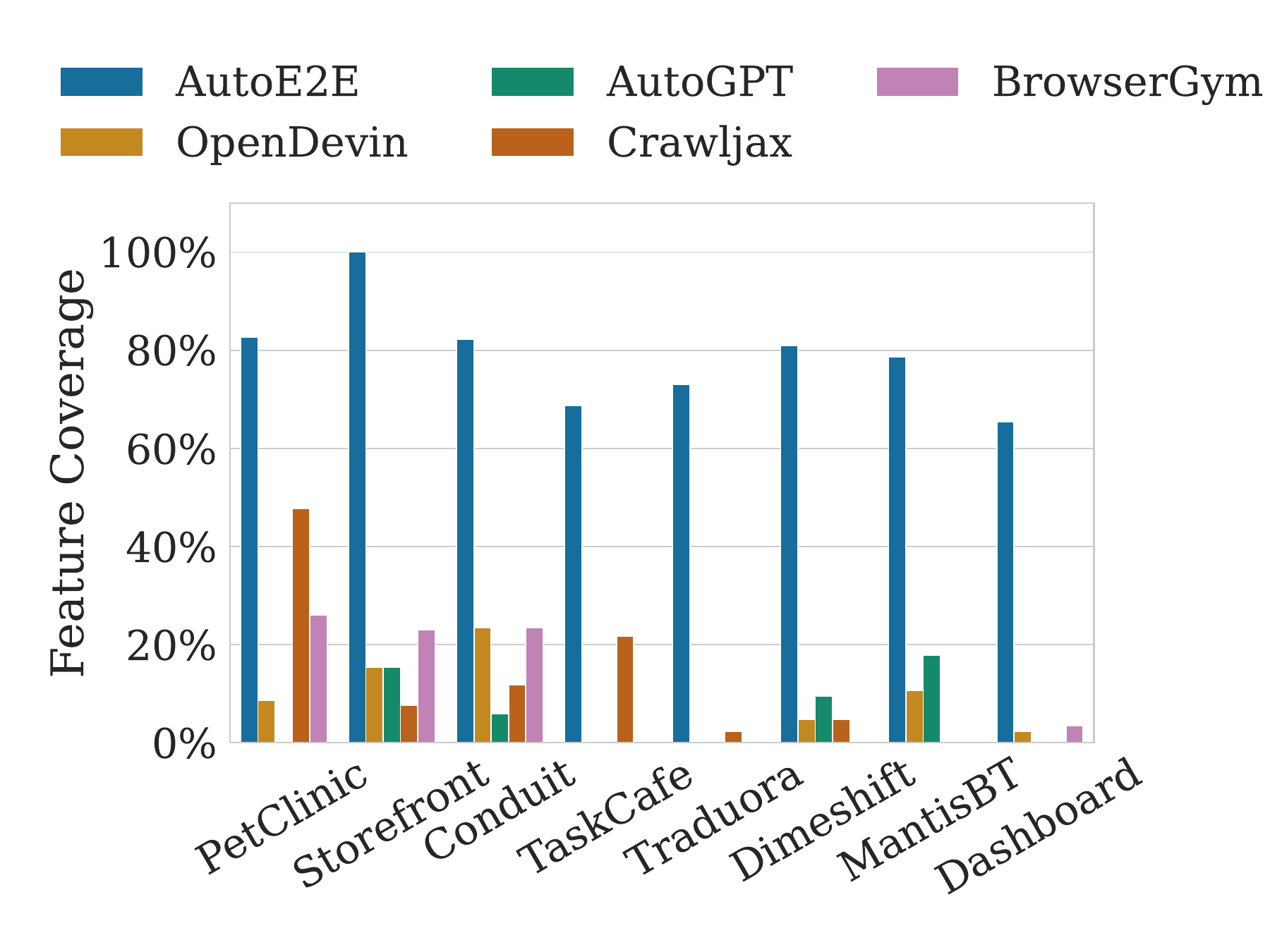}
    \caption{Feature Coverage of different methods on the subjects}
    \label{fig:method-coverage}
\end{figure}

We evaluated the generated tests by each baseline against \benchname as described in \autoref{sec:benchmark}. The Feature Coverage results for different subjects and baselines are presented in \autoref{fig:method-coverage}. \toolname achieves an average Feature Coverage rate of \ourscoverage, significantly outperforming \crawljax (\crawljaxcoverage), \webcanvas (\webcanvascoverage), \browsergym (\browsergymcoverage), \autogpt (\autogptcoverage), and \opendevin (\opendevincoverage). Notably, \toolname surpasses the next best performing tool, \crawljax, by \oursimprovement, and best agent-based tool, \browsergym, by \oursimprovementagent. \webcanvas has not been included in \autoref{fig:method-coverage} since it did not generate any test cases.

The F1 scores for the generated test cases were \crawljaxf for \crawljax, \browsergymf for \browsergym, \autogptf for \autogpt, and \opendevinf for \opendevin, compared to \oursf for \toolname.

\subsubsection{Feature Complexity}
While the coverage rate treats all features equally, it is important to acknowledge that features vary in complexity. Some features within an application necessitate longer chains of ordered actions to be successfully executed.  Different tools may encounter difficulties with extended action chains due to the challenge of maintaining context over longer periods. \autoref{tab:feature-complexity} provides a statistical analysis of action chain complexity across different tools.

\begin{table}[h]    
\centering
{\small
    \caption{Feature action chain length}
    \label{tab:feature-complexity}
    \begin{tabular}{lrrrr}
        \toprule
        \textbf{Tool Name} & \textbf{Min} & \textbf{Max} & \textbf{Average} & \textbf{Median} \\
        \midrule

        \rowcolor{lightgray}
        \toolname & 1 & 7 & \ourschain & 4 \\

        \crawljax & 1 & 7 & \crawljaxchain & 3 \\
        
        \rowcolor{lightgray}
        \webcanvas & 0 & 0 & 0.0 & 0 \\

        \browsergym & 1 & 3 & 2.0 & 2 \\

        \rowcolor{lightgray}
        \autogpt & 1 & 3 & \autogptchain & 1 \\

        \opendevin & 1 & 4 & \opendevinchain & 1 \\

        \rowcolor{lightgray}
        All Features & 1 & 7 & \totalchain & 4 \\
        \bottomrule
    \end{tabular}
}
\end{table}

The average length of feature chains in the benchmark is \totalchain actions. Notably, \toolname demonstrates proficiency in handling longer chains, averaging \ourschain actions per feature, compared to \crawljaxchain for \crawljax, \browsergymchain for \browsergym, \autogptchain for \autogpt, and \opendevinchain for \opendevin.

\subsubsection{Test Case Count}
The number of test cases generated varies across different tools, as each tool employs its own criteria for determining the appropriate quantity. This raises the question of whether forcing a tool to generate more test cases would necessarily lead to increased feature coverage. The lower half of \autoref{fig:rank-coverage} sheds light on this by illustrating the moving average of Feature Coverage for the methods across all subjects. Notably, the baselines reach a plateau in coverage at a relatively low rate, suggesting that simply increasing the number of test cases does not guarantee a corresponding increase in feature coverage.

\section{Discussion}
The framework developed in this work, particularly the findings from Section \ref{sec:modeling}, is platform- and implementation-agnostic. This means the underlying principles can be extended to generate test cases for other platforms, such as mobile applications. Furthermore, this implementation independence allows for significant potential improvements in performance and cost-effectiveness in future iterations, as the framework is not rigidly tied to any specific technology or tool. Additionally, the introduction of \benchname provides a standardized and automated means of evaluating E2E test case generation techniques. This benchmark fills a crucial gap in the research community, enabling more rigorous comparison and development of novel approaches in the field of automated E2E test generation.



\header{Limitations}
Despite its strengths, our approach has limitations. Currently, \toolname focuses on a one-to-one mapping between test cases and features, whereas real-world scenarios often require multiple test cases per feature to assess diverse interactions. Additionally, our implementation generates assertions using the entire state after each action, which may not always be the most meaningful or targeted approach.


\header{Threats to Validity}
It is important to acknowledge potential threats to the validity of our findings and the steps taken to mitigate them. One such threat lies in the representativeness of the web applications selected for \benchname. To address this, we have carefully curated a diverse set of applications spanning a wide range of categories. This diversity aims to enhance the generalizability of our results and reduce potential biases associated with specific application types.

Another potential threat stems from the inherent subjectivity in defining and extracting features. To mitigate this, we have established a precise and formal definition of a feature (Definition \ref{def:application-feature}). Furthermore, we employed multiple authors in the feature identification process during benchmark construction, promoting a more objective and comprehensive perspective.

Finally, the use of different LLMs in the evaluation process could pose a threat to validity. While our goal was to maintain consistency by utilizing the same LLMs across all methods, certain constraints necessitated variations. Specifically, among the selected agents and baselines, only \opendevin supported \sonnet, while others (\autogpt, \webcanvas, \browsergym) lacked support for Anthropic models at the time of evaluation, requiring the use of \gpt. However, it is important to note that both \sonnet and \gpt exhibit very similar performance in reported benchmarks~\cite{gpt4oresults, claude3results}. Therefore, we anticipate that these variations in LLM usage would not significantly impact the overall performance trends observed in our evaluation.

\section{Related Work}
\header{Web Navigation Agents}
In the academic context, extensive research has been conducted on automating the execution of tasks defined in natural language~\cite{humphreys2022data, li2021glider, mazumder2020flin, jia2019dom, xu2021grounding, gur2018learning, liu2018reinforcement}. The aim is for an agent to determine and execute a sequence of actions within a web application that fulfills these instructions.
This research is divided into two principal categories: traditional methods utilizing Reinforcement Learning (RL) agents and newer approaches that focus on LLMs.

The traditional techniques involve deep learning models to translate natural language instructions into embeddings~\cite{humphreys2022data, li2021glider, mazumder2020flin, jia2019dom, gur2018learning, liu2018reinforcement}. These embeddings are then mapped to specific actions on the web page. Variations in these methods are seen in the architecture of the deep learning models, the strategies of the RL policies, and how actions are modeled. Additionally, some methods might include or omit certain components. Techniques in this category often incorporate demonstrations by expert users~\cite{gur2018learning, liu2018reinforcement}, use heuristics based on human-designed systems~\cite{zheng2021automatic}, or confine the agent's actions to a predetermined set~\cite{jia2019dom, xu2021grounding}.

Contrastingly, newer methods leveraging LLMs typically forgo the RL training phase. These models delegate the decision-making about subsequent actions to the LLMs~\cite{wang2023survey, gur2022understanding, furuta2023multimodal, sodhi2023heap}. These newer agents have been proposed by academic research, open-source communities ~\cite{langchain, autogpt2024}, and large-scale companies~\cite{achiam2023gpt}.

\header{LLM-based Testing}
Recent studies have increasingly focused on using LLMs to automate software and web testing processes. Some research concentrates on software unit test generation \cite{llmunittest, chen2024chatunitest, chatgptunittest}. Some studies focus on accessibility testing, utilizing LLMs to identify and address accessibility issues \cite{axnav}. Other studies have directed their attention towards GUI testing, including software GUI testing \cite{guisoftwaretesting} and mobile application GUI testing \cite{liu2024make, liu2024testing, fillintheblank}. Our recent work has focused on automated web form testing, where LLMs simulate user interactions and validate form functionalities through constraint-based testing~\cite{alian2024bridging}.

\header{E2E Test Generation}
A specific but less extensive area of research focuses on generating end-to-end (e2e) test scenarios. Earlier approaches in this field worked on mapping applications to page objects and creating test cases for them~\cite{stoccoapogen}. However, recent studies use LLMs for automated test generation~\cite{wang2024xuat, liu2305chatting, liu2023make}. For example, a recent prominent study~\cite{liu2023make} uses prompting patterns to track executed actions, directing the LLM to select actions based on the high-level task in the application. These actions are then formulated into test cases for the overarching task.


\section{Conclusion}
Automated E2E test case generation remains challenging. In this paper, we formally defined the problem, introduced a novel methodology (\toolname) for feature-driven test generation, and developed \benchname, a benchmark for automated evaluation. \toolname achieves \ourscoverage feature coverage, outperforming baselines by \oursimprovement. Future work includes enhancing \toolname's performance and exploring assertion generation for more comprehensive test coverage.

\section{Data Availability}
We have made \toolname and \benchname publicly available \cite{autoe2e} to facilitate reproducibility of our results. Detailed instructions for replicating our experimental setup are also provided.

\bibliographystyle{IEEEtran}
\interlinepenalty=10000
\bibliography{references}

\end{document}